\documentclass[pdflatex,sn-mathphys-num]{sn-jnl}% Math and Physical Sciences Numbered Reference Style
\usepackage{graphicx}%
\usepackage{multirow}%
\usepackage{amsmath,amssymb,amsfonts}%
\usepackage{amsthm}%
\usepackage{mathrsfs}%
\usepackage[title]{appendix}%
\usepackage{xcolor}%
\usepackage{textcomp}%
\usepackage{manyfoot}%
\usepackage{booktabs}%
\usepackage{algorithm}%
\usepackage{algorithmicx}%
\usepackage{algpseudocode}%
\usepackage{listings}%

\usepackage{pdfpages}

\theoremstyle{thmstyleone}%
\theoremstyle{thmstyletwo}%

\theoremstyle{thmstylethree}%

\begin{document}

\title[Article Title]{Hybrid antiferroelectric--ferroelectric domain walls\\ in noncollinear antipolar oxides}


\author[1]{\fnm{Ivan N.} \sur{Ushakov}}%\email{ivan.ushakov@ntnu.no}

\author[2]{\fnm{Mats} \sur{Topstad}}

\author[1]{\fnm{Muhammad Z.} \sur{Khalid}}

\author[3]{\fnm{Niyorjyoti} \sur{Sharma}}

\author[4]{\fnm{Christoph} \sur{Grams}}

\author[1]{\fnm{Ursula} \sur{Ludacka}}

\author[1]{\fnm{Jiali} \sur{He}}

\author[1,5]{\fnm{Kasper A.} \sur{Hunnestad}}

\author[1]{\fnm{Mohsen} \sur{Sadeqi-Moqadam}}

\author[1]{\fnm{Julia} \sur{Glaum}}

\author[1]{\fnm{Sverre M.} \sur{Selbach}}

\author[4]{\fnm{Joachim} \sur{Hemberger}}

\author[6]{\fnm{Petra} \sur{Becker}}

\author[6]{\fnm{Ladislav} \sur{Bohatý}}

\author[3]{\fnm{Amit} \sur{Kumar}}

\author[7,8]{\fnm{Jorge} \sur{Íñiguez-González}}%\email{jorge.iniguez@list.lu}

\author[2]{\fnm{Antonius T. J.} \sur{van Helvoort}}

\author*[1]{\fnm{Dennis} \sur{Meier}}\email{dennis.meier@ntnu.no}

\affil[1]{\orgdiv{Department of Materials Science and Engineering}, \orgname{Norwegian University of Science and Technology (NTNU)}, \orgaddress{\city{Trondheim}, \country{Norway}}}

\affil[2]{\orgdiv{Department of Physics}, \orgname{Norwegian University of Science and Technology (NTNU)}, \orgaddress{\city{Trondheim}, \country{Norway}}}

\affil[3]{\orgdiv{Centre for Quantum Materials and Technologies (CQMT)}, \orgname{Queen's University Belfast}, \orgaddress{\city{Belfast}, \country{United Kingdom}}}

\affil[4]{\orgdiv{Institute of Physics II}, \orgname{University of Cologne}, \orgaddress{\city{Cologne}, \country{Germany}}}

\affil[5]{\orgdiv{Department of Electronic Systems}, \orgname{Norwegian University of Science and Technology (NTNU)}, \orgaddress{\city{Trondheim}, \country{Norway}}}

\affil[6]{\orgdiv{Institute of Geology and Mineralogy}, \orgname{University of Cologne}, \orgaddress{\city{Cologne}, \country{Germany}}}

\affil[7]{\orgdiv{Smart Materials Unit}, \orgname{Luxembourg Institute of Science and Technology (LIST)}, \orgaddress{\city{Esch/Alzette}, \country{Luxembourg}}}

\affil[8]{\orgdiv{Department of Physics and Materials Science}, \orgname{University of Luxembourg}, \orgaddress{\city{Belvaux}, \country{Luxembourg}}}

%\affil[3]{\orgdiv{Department}, \orgname{Organization}, \orgaddress{\street{Street}, \city{City}, \postcode{610101}, \state{State}, \country{Country}}}


\abstract{Antiferroelectrics are emerging as advanced functional materials and are fertile ground for unusual electric effects. For example, they enhance the recoverable energy density in energy storage applications and give rise to large electromechanical responses. Here, we demonstrate noncollinearity in dipolar order as an additional degree of freedom, unlocking physical properties that are symmetry-forbidden in classical antiferroelectrics. We show that noncollinear order of electric dipole moments in K$_3$[Nb$_3$O$_6|$(BO$_3$)$_2$] leads to a coexistence of antiferroelectric and ferroelectric behaviors. Besides the double-hysteresis loop observed in antiferroelectrics, a pronounced piezoresponse and electrically switchable domains are observed, separated by atomically sharp and micrometer-long charged domain walls. Hybrid ferroelectric--antiferroelectric responses are expected in a wide range of noncollinear systems, giving a new dimension to the research on antiferroelectrics and multifunctional oxides in general.}




\maketitle

\section*{Introduction}\label{sec1}

In 1928, Louis Néel introduced a new way of looking at magnetism~\cite{Neel1972}, establishing the fundamentals of antiferromagnetism. About 20 years later, Charles Kittel expanded the concept towards antiferroelectrics, defining them as systems with lines of spontaneously polarized ions pointing in antiparallel directions, so that the macroscopic polarization is zero (Fig.~\ref{fig:figure_1}(a))~\cite{Kittel1951}. Due to the absence of a macroscopic magnetization or polarization, such antiferroics were initially considered to be of limited technological interest compared to their ferroic counterparts. This perception has changed completely and today, antiferroics, such as antiferromagnets and antiferroelectrics, are intensively studied as key functional materials for next-generation spintronics~\cite{Jungwirth2016}\cite{Han2023} and energy applications~\cite{Liu2018}\cite{Randall2021}, respectively.

Going beyond collinear antiferroelectrics, it was discussed in the 1970s that antipolar dipole structures are compatible with polar point groups and hence, in principle, also allow for noncollinear antipolar order as sketched in Fig.~\ref{fig:figure_1}(b)~\cite{Blinc1974}\cite{Zheludev1971}\cite{Scott1974}. Examples of bulk systems with noncollinear antipolar order include Gd$_2$(MoO$_4$)$_3$~\cite{Jeitschko1972}, BiCu$_x$Mn$_{7-x}$O$_{12}$~\cite{Khalyavin2020}, (Pb$_{0.91}$La$_{0.06}$)(Zr$_{0.42}$Sn$_{0.40}$Ti$_{0.18}$)O$_3$~\cite{Gao2022}, and some organic salts~\cite{Wang2024}. A theoretical discussion on Dzyaloshinskii–Moriya-like interactions in antiferroelectrics provided additional insights, emphasizing the possibility to enhance their functionality by introducing noncollinearity~\cite{Zhao2021}. In general, analogous to noncollinear spin textures, a canting of electric dipole moments reduces the symmetry, which can be leveraged to unlock otherwise symmetry-forbidden properties. In contrast to magnetic materials, however, the concept of noncollinearity in electric systems is much less explored and emergent physical phenomena beyond those observed in classical antiferroelectrics remain to be demonstrated. 

In this study, we investigate single crystals of potassium niobate borate K$_3$[Nb$_3$O$_6|$(BO$_3$)$_2$], which were grown from the melt as described elsewhere \cite{Becker1996}. While the crystals are hexagonal at growth temperature (space group P$\bar{6}2$m, 189) \cite{Becker1997}, they possess orthorhombic symmetry at room temperature (space group P$2_1$ma, 26) \cite{Becker1996}, with a quadruplicated unit cell volume. This symmetry relation belongs to the Aizu species $\bar{6}2$mFm2m and leads to the occurrence of three polar 120° ferroelastic domain states in the orthorhombic phase (labeled T$'$, T$''$ and T$'''$ in the following) \cite{Aizu1970}\cite{Erb2020}\cite{Authier2014}. From the point of view of symmetry, ferroelectricity is permitted and both neutral and charged domain walls may arise between the domains \cite{Sapriel1975}\cite{Erhart2004}, which motivates our choice of K$_3$[Nb$_3$O$_6|$(BO$_3$)$_2$] as model system. At the micrometer scale, the ferro-elastic domains can readily be visualized in a polarization microscope between crossed polarizers \cite{Kaminskii2004}, as seen in Fig.~\ref{fig:figure_1}(c), whereas their nanoscale physics remain to be explored.

\section*{Results}\label{sec2}

\subsection*{Noncollinear antiferroelectric order and macroscopic switching behavior}

\begin{figure}
    \centering
    \includegraphics[width=1\textwidth]{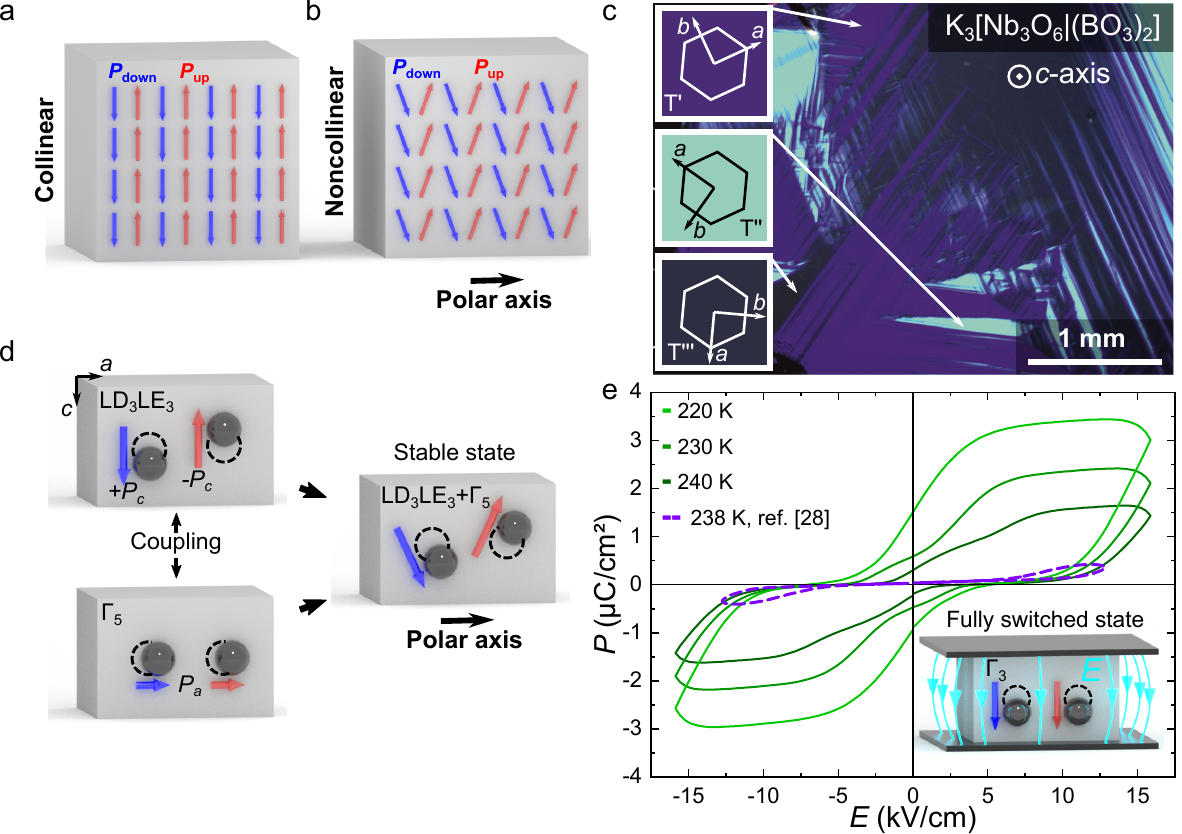}
    \caption{\textbf{Noncollinear antipolar order in K$_3$[Nb$_3$O$_6|$(BO$_3$)$_2$].} 
    \textbf{a}, In conventional antiferroelectrics, the electric dipole moments (blue and red arrows) are antiparallel, which gives zero net polarization and conserves inversion symmetry. \textbf{b}, A canting of the antipolar order reduces the symmetry and leads to the formation of a polar axis, enabling additional physical properties that are symmetry-forbidden in collinear systems. \textbf{c}, Polarized light microscopy of the K$_3$[Nb$_3$O$_6|$(BO$_3$)$_2$] (001) sample face reveals three ferroelastic $120^\circ$ twin domain states, labeled as $\mathrm{T}'$, $\mathrm{T}''$, and $\mathrm{T}'''$. The respective crystallographic orientations are sketched as insets with the $a$-axis being the polar axis. \textbf{d}, In K$_3$[Nb$_3$O$_6|$(BO$_3$)$_2$], noncollinear antipolar order arises from the coupling of antipolar (LD$_3$LE$_3$) and polar ($\Gamma_5$) modes, defining the system as a proper antiferroelectric, improper ferroelectric (and ferroelastic) material. \textbf{e}, For electric fields, $E$ (15~mHz), applied along the $c$-axis (perpendicular to the polar axis), a double hysteresis loop opens up for $P(E)$ as the temperature decreases, consistent with previous experiments conducted at higher frequencies (purple dashed line, adapted from ref.~\cite{Shan2020}). The switching behavior is due to a polar instability ($\Gamma_3$) parallel to the antipolar direction as sketched in the inset, leading to an electric-field induced polar state.
    \label{fig:figure_1}
    }
\end{figure}

 We begin by analyzing the microscopic origin of the dipolar order in K$_3$[Nb$_3$O$_6|$(BO$_3$)$_2$]. Our density functional theory (DFT) calculations reveal an antipolar instability of the high-temperature phase, with local electric dipoles forming along the $c$-axis (LD$_3$LE$_3$ mode), as illustrated in Fig.~\ref{fig:figure_1}(d). By performing a customary symmetry analysis using ISODISTORT \cite{Campbell2006} and INVARIANTS \cite{Hatch2003}, we conclude that the LD$_3$LE$_3$ mode is the primary order parameter driving the phase transition to the room-temperature phase. Because of the threefold symmetry in the hexagonal high-temperature phase, however, a pure anti-polar arrangement cannot be geometrically satisfied (Supplementary Fig. S1). The LD$_3$LE$_3$ mode breaks the threefold symmetry, introducing a polar axis. In addition, we find a secondary mode of polar character and symmetry $\Gamma_5$, that acts as an improper order and leads to ferroelectricity (Fig. 1(d)). Thus, the room-temperature P$2_1$ma phase -- where both the LD$_3$LE$_3$ and $\Gamma_5$ modes coexist -- can be characterized as proper antiferroelectric and improper ferroelectric, with a noncollinear (canted) antipolar order. Our calculations give net polarization components $P_a~\approx 0.03$~$\mu$C/cm$^2$ and $P_c~\approx \pm 19$~$\mu$C/cm$^2$. For reference, the resulting spontaneous polarization, $P_a$, is about 500 times smaller than in BaTiO$_3$ ($\approx 15$~$\mu$C/cm$^2$)\cite{Merz1949} and one to two orders of magnitude smaller than in other improper ferroelectrics, such as Cu$_3$B$_7$O$_{13}$Cl ($\approx~1.8$~$\mu$C/cm$^2$)~\cite{Schmid1980} and Gd$_2$(MoO$_4$)$_3$ ($\approx~0.2$~$\mu$C/cm$^2$)~\cite{Cummins1970}.

 Importantly, because of the antipolar components $\pm P_c$, K$_3$[Nb$_3$O$_6|$(BO$_3$)$_2$] displays a double-hysteresis loop for electric fields applied along the $c$-axis~\cite{Shan2020}, i.e., perpendicular to the polar $a$-axis. Such double-hysteresis loops are commonly considered a hallmark of antiferroelectricity~\cite{Randall2021}\cite{Park1997}\cite{Fesenko1978}. Figure~\ref{fig:figure_1}(e) shows respective low-frequency (15 mHz) $P_c(E)$ measurements (Method section) recorded on our sample together with literature data (50 Hz)~\cite{Shan2020}. Within the electric-field range of our experiment, we observed non-saturated double-hysteresis loops with a maximum electric-field-induced polarization along the antipolar $c$-axis of about $3.5$~$\mu$C/cm$^2$ ($T=220$~K).

 Based on our symmetry analysis, we attribute the switching behavior to the coexistence of the primary antipolar (LD$_3$LE$_3$) and a secondary polar ($\Gamma_5$) mode ($E=0$). A field-aligned state ($\Gamma_3$) is approached for electric fields applied parallel ($+E$) or antiparallel ($-E$) to the antipolar $c$-axis as sketched in the inset to Fig. \ref{fig:figure_1}(e), reverting to noncollinear antipolar order as the electric field is removed.

%The Introduction section, of referenced text \cite{bib1} expands on the background of the work (some overlap with the Abstract is acceptable). The introduction should not include subheadings.

%Springer Nature does not impose a strict layout as standard however authors are advised to check the individual requirements for the journal they are planning to submit to as there may be journal-level preferences. When preparing your text please also be aware that some stylistic choices are not supported in full text XML (publication version), including coloured font. These will not be replicated in the typeset article if it is accepted. 

\subsection*{Hybrid antiferroelectric--ferroelectric domain walls}

\begin{figure}
    \centering
    \includegraphics[width=0.5\textwidth]{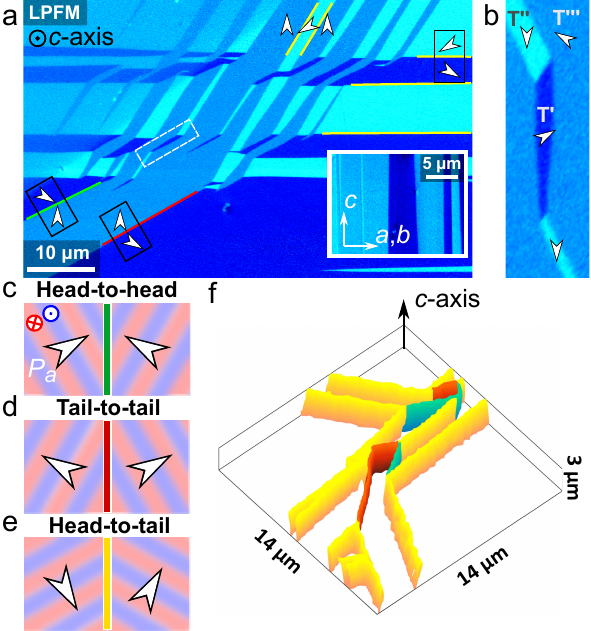}
    \caption{\textbf{Coexistence of neutral and charged domain walls.}
    \textbf{a}, Lateral PFM on a (001) sample face, plotted as $A\cos{\phi}$, where $A$ is the amplitude an $\phi$ is the phase. The inset shows lateral PFM of a surface with the $c$-axis in-plane. Arrows indicate the strongest piezoresponse directions in the three types of domains (T$'$, T$''$, and T$'''$) as labeled in (\textbf{b}). This direction coincides with the direction of $P_a$. Three domain wall types are observed: (\textbf{c}) positively charged head-to-head (green), (\textbf{d}) negatively charged tail-to-tail (red), and (\textbf{e}) neutral head-to-tail (yellow) walls. The coexisting antipolar order ($\pm P_c$) is indicated by pink ($-P_c$) and violet $+P_c$ colors. \textbf{f}, Tomographic atomic force microscopy data showing the 3D structure of the different types of domain walls.
    \label{fig:figure_2}
    }
\end{figure}

We expand our study by investigating the domains in K$_3$[Nb$_3$O$_6|$(BO$_3$)$_2$] towards the so far unexplored nanoscale. Figure~\ref{fig:figure_2}(a) shows a piezoresponse force microscopy (PFM) image (lateral signal, LPFM) recorded on a surface perpendicular to the $c$-axis. For the PFM measurement, an a.c.\ voltage of 5~V and 40.13 kHz is applied to a diamond coated conductive tip. Prior to the experiment, the sample was chemo-mechanically polished with silica slurry, giving a flat surface with root-mean-square roughness of 696$\pm$0.9~pm. The PFM image shows a pronounced piezoresponse with three contrast levels. We note that this behavior is fundamentally different from classical collinear antiferroelectrics, such as prototypical PbZrO$_3$, where piezoelecricity in the antiferroelectric state is forbidden by symmetry~\cite{Shirane1951}\cite{Teslic1998}.

By performing vector PFM (Supplementary Note 2), we determine the orientation of the polar $a$-axis within the different domains. This orientation is identical to the direction in which the piezoresponse has its maximum (Supplementary Note 1) and gives the direction of $P_a$ as indicated by the white arrows in Fig.~\ref{fig:figure_2}(a). We identify three possible domain states with $P_a$ changing by 120$^{\circ}$ across the domain walls, consistent with the expected ferroelastic twin domain states T$'$, T$''$, and T$'''$, and the stripe-like domain structure observed at larger length scales (Fig.~\ref{fig:figure_1}(c)). The PFM data reveals, however, that the domain structure is more complex than suggested by the optical experiments. At the nanoscale, the system exhibits chevron-like patterns with sequences of alternating domain states, some ending in narrow needle-shaped domains as presented in Fig. \ref{fig:figure_2}(b). These structures coexist with the optically resolved larger domains, resulting in an intriguing pattern of intertwined micro- and nanometer-sized domains. The behavior is classical to ferroelastics, allowing the material to minimize mechanical strain \cite{Salje1991}\cite{Sui2020}.

On a closer inspection of Fig.~\ref{fig:figure_2}(a), we find that the domain walls display different polarization configurations with respect to $P_a$, including head-to-head, tail-to-tail, and head-to-tail arrangements, as sketched and color-labeled in Fig.~\ref{fig:figure_2}(c)-(e). PFM images gained on a perpendicular surface (inset in Fig.~\ref{fig:figure_2}(a)) and PFM tomography (Fig.~\ref{fig:figure_2}(f) and Supplementary Note 2) reveal that all domain walls propagate parallel to the $c$-axis, forming planar 2D systems. As expected from the symmetry considerations given above, the majority of domain walls is in head-to-tail configuration, although head-to-head and tail-to-tail domain walls are also frequently observed. Similar to the head-to-tail walls, these walls extend over several micro\-meters. The latter is remarkable, because head-to-head and tail-to-tail walls represent polar discontinuities that carry bound surface charges $\pm 2P_a \cos 30^{\circ} = \pm 0.05$~$\mu$C/cm$^2$, which makes them energetically costly. So far, charged planar domain walls with comparable physical dimensions have only been observed in boracites~\cite{Mcquaid2017}. In case of the boracites, however, their injection requires the application of point pressure, whereas the charged head-to-head and tail-to-tail walls naturally form in K$_3$[Nb$_3$O$_6|$(BO$_3$)$_2$].

It is important to note that the domain walls resolved in Fig.~\ref{fig:figure_2} are fundamentally different from previously studied interfaces in both ferroelectrics and antiferroelectrics as the spontaneous polarization $P_a$ and the anti-polar component $\pm P_c$ simultaneously change across the walls as illustrated in Fig.~\ref{fig:figure_2}(c)-(e), which classifies them as hybrid antiferroelectric-ferroelectric (and ferroelastic) domain walls.

\begin{figure}
    \centering
    \includegraphics[width=0.5\textwidth]{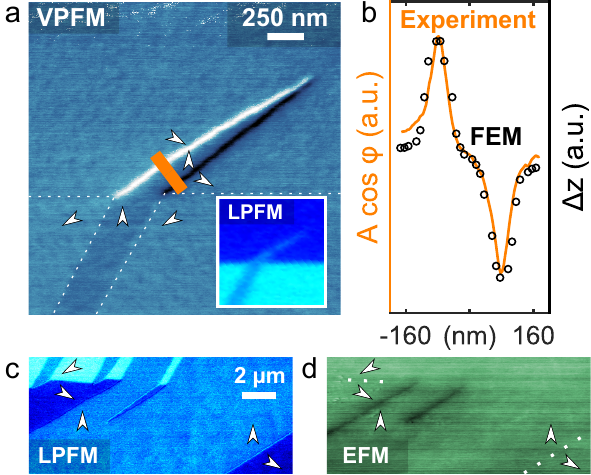}
    \caption{\textbf{Electomechanical and electrical domain wall properties.}
    \textbf{a}, Vertical resonance PFM on a (001) sample face (plotted as $A\cos{\phi}$) shows a pronounced piezoelectric response at the head-to-head and tail-to-tail domain walls. Corresponding LPFM data is presented in the inset to (\textbf{a}). \textbf{b}, Profile of the head-to-head and tail-to-tail domain walls in (\textbf{a}) and FEM simulation of the static displacement ($\Delta z$) in the $c$-direction. The comparison of (\textbf{c}) LPFM ($A \cos \phi$) and (\textbf{d}) EFM (amplitude, (001) face) scans shows that the strongest electrostatic response arises at the head-to-head domain walls. 
    }
    \label{fig:figure_3}
\end{figure}

Next, we analyze the local electronic properties at these hybrid domain walls. We note that up to voltages of $\pm$150~V, K$_3$[Nb$_3$O$_6|$(BO$_3$)$_2$] behaves insulating in conductive atomic force microscopy measurements with no detectable differences between domains and domain walls (not shown). The high resistance measured at the local scale is consistent with the large band gap of the material (colorless crystals). We thus focus on the local electromechanical and electrostatic properties in Fig.~\ref{fig:figure_3}. Figure~\ref{fig:figure_3}(a) shows a high-resolution PFM image (vertical signal, VPFM) of one of the needle-shaped domains shown in the inset. A pronounced electromechanical response is measured at the head-to-head (bright) and tail-to-tail (dark) domain walls, which clearly distinguishes them from the rest of the material, that is, the twin domains and the neutral head-to-tail domain walls.

To understand the anomalous local response, we perform finite elements simulations (FEM, Supplementary Note 1). Consistent with recent studies of the piezoresponse at charged ferroelectric domain walls~\cite{Lu2023}, the model shows that the shear strain changes sign at the hybrid antiferroelectric-ferroelectric domain walls, leading to an upward displacement (bright) at the head-to-head domain walls and a downward displacement (dark) at the tail-to-tail domain walls. In Figure~\ref{fig:figure_3}(b), we present the simulated contraction/expansion for neighboring head-to-head and tail-to-tail domain walls. The simulation result is in excellent agreement with the cross-sectional VPFM data recorded along the orange line in Fig.~\ref{fig:figure_3}(a) and fully consistent with the direction of the polar axis determined by vector PFM. We thus conclude that the distinct electromechanical response at the domain walls originates from counteracting shear strains, enabled by the primary LD$_3$LE$_3$ mode.

The electrostatic properties associated with the hybrid antiferroelectric-ferroelectric domain walls are presented in Fig.~\ref{fig:figure_3}(c,d). By correlating LPFM and electrostatic force microscopy (EFM, Method section) data, we find that the positively charged head-to-head domain walls exhibit an electrostatic response different from the domains, being visible as straight dark lines in the EFM map (Fig.~\ref{fig:figure_3}(d)). In contrast, no signature is resolved in this EFM scan at the negatively charged tail-to-tail and the neutral head-to-tail domain walls. We note that occasionally a weak EFM signal is also detected at tail-to-tail and head-to-tail domain walls. These signals, however, are typically more than 4-5 times smaller than for the head-to-head domain walls and at the resolution limit of our experiment (Supplementary Fig. S2). We also note that the surface potential variations do not come from topographical contributions (Supplementary Fig. S3). The observation that the surface potential varies mainly at the position of the positively charged head-to-head walls suggests that the material either efficiently screens the negative bound charges at the tail-to-tail domain walls (e.g., by accumulation of mobile hole carriers)~\cite{Meier2012}\cite{Eliseev2011}\cite{Gureev2011} or it develops a non-trivial domain wall morphology to avoid the emergence of a polar discontinuity~\cite{Tikhonov2022}.

\subsection*{Atomic-scale domain wall structure}

\begin{figure}
    \centering
    \includegraphics[width=1\textwidth]{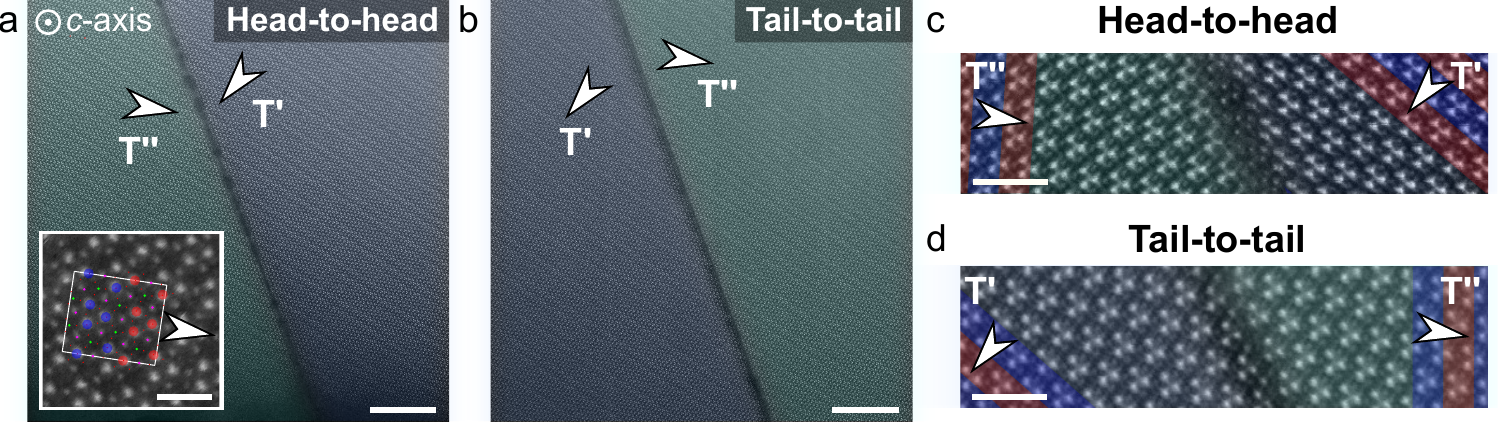}
    \caption{\textbf{Atomic structure of hybrid antiferroelectric-ferroelectric domain walls.}
    \textbf{a}, False-colored HAADF-STEM image of the (001) sample face showing a head-to-head domain wall. The scalebar is 15 nm. The inset shows a high-resolution image of the Nb atoms (highlighted red and blue) gained on a different FIB-prepared sample. The scalebar is 1 nm.
    \textbf{b}, Same as in \textbf{a} for a tail-to-tail domain wall. The scalebar is 15 nm.
    \textbf{c}, High-resolution HAADF-STEM image of a head-to-head domain wall. False colors (red and blue) illustrative how the antipolar order changes across the domain wall. The scalebar is 2 nm. \textbf{d}, Same as in \textbf{c} for a tail-to-tail domain wall. The scalebar is 2 nm.
    \label{fig:figure_4}
    }
\end{figure}

High-resolution images of the atomic-scale structure at head-to-head and tail-to-tail domain walls are presented in Fig.~\ref{fig:figure_4}. The high-angle annular dark-ﬁeld scanning transmission electron microscopy (HAADF-STEM) images are recorded viewing along the $c$-axis of K$_3$[Nb$_3$O$_6|$(BO$_3$)$_2$] with the bright dots representing the Nb atomic columns as illustrated in the inset to Fig.~\ref{fig:figure_4}(a). The domain walls are visible as straight dark lines in Fig.~\ref{fig:figure_4}(a,b), whereas a perfectly ordered orthorhombic lattice is observed away from the walls. Across the walls, the polar axis rotates by 120$^{\circ}$, forming nominally charged head-to-head (Fig.~\ref{fig:figure_4}(a)) and tail-to-tail (Fig.~\ref{fig:figure_4}(b)) configurations, respectively, as confirmed by correlated PLM-PFM-TEM measurements (Supplementary Fig. S4). Importantly, the direction in which the antipolar displacement of Nb atoms is modulated changes together with the polar axis from one domain to the next as indicated by the white arrows in Fig.~\ref{fig:figure_4}(a,b). The change in the antipolar order is highlighted in Fig.~\ref{fig:figure_4}(c,d), where blue and red colors indicate polarization up ($+P_c$) and down ($-P_c$), respectively.

The HAADF-STEM data shows the hybrid nature of the domain walls at the atomic scale, exhibiting characteristic discontinuities in both the antiferroelectric and ferroelectric displacement patterns. Independent of their charge state, the hybrid antiferroelectric-ferroelectric walls form atomically sharp interfaces with a lateral extension of several micrometers (Fig.~\ref{fig:figure_2}) and a width in the sub-unit-cell regime (Fig.~\ref{fig:figure_4}). This width is comparable to the size of the Nb trimers in K$_3$[Nb$_3$O$_6|$(BO$_3$)$_2$]. Across all length scales investigated in our study, we thus observe that the domain walls are almost perfectly planar, independent of their actual charge state.

\subsection*{Electric-field control of hybrid domain walls}

\begin{figure}
    \centering
    \includegraphics[width=0.5\textwidth]{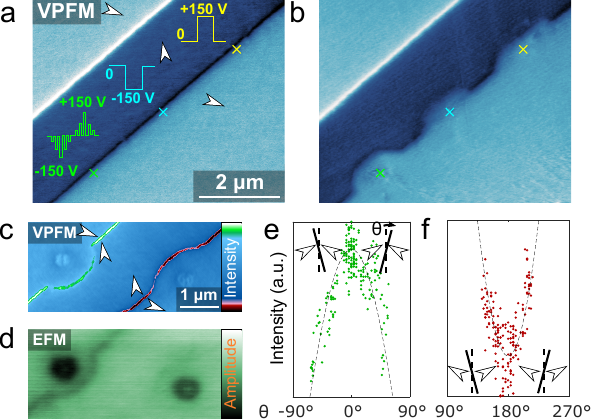}
    \caption{\textbf{Local switching behavior at head-to-head and tail-to-tail domain walls.} \textbf{a}, Localized electric d.c.\ and a.c.\ fields of up to $\pm 150$~V are generated by the probe tip and applied close to the tail-to-tail wall at the positions marked by green (a.c.\ $\pm 150$~V), cyan (d.c.\ $-150$~V), and yellow (d.c.\ $+150$~V) crosses. \textbf{b}, Independent of the polarity of the applied voltage, the tail-to-tail wall moves in the direction of the neighboring head-to-head wall. \textbf{c}, Vice versa, head-to-head walls move in the direction of the nearest tail-to-tail wall when an electric field is applied. \textbf{d} EFM data collected at the same position as the VPFM scan in (\textbf{c}) after application of the electric field demonstrates that the domain-wall related electrostatic anomaly (dark contrast) moves together with the domain wall. \textbf{e}, The analysis of the VPFM signal in the curved section of the head-to-head wall in (\textbf{c}) reveals that the piezoresponse varies along with the domain wall orientation, showing as sinusoidal relationship between the magnitude of piezoresponse and the orientation angle $\theta$. \textbf{f}, Same as in (\textbf{e}) for the tail-to-tail wall.} 
    \label{fig:figure_5}
\end{figure}

To explore the general possibility to control the position of the hybrid antiferroelectric-ferroelectric domain walls, we investigate their response to local electric field poling (Fig.~\ref{fig:figure_5}). Figure~\ref{fig:figure_5}(a,b) demonstrates that the position of individual domain walls can readily be controlled by application of an electric field, exemplified for the case of a tail-to-tail domain wall. PFM images are recorded before (Fig.~\ref{fig:figure_5}(a)) and after (Fig.~\ref{fig:figure_5}(b)) applying a d.c.\ or a.c.\ bias voltage (see Supplementary Fig. S5) to the probe tip, which rested at the positions marked by the crosses (green: $\pm 150$~V, cyan: -150~V, yellow: +150~V). We observe that, independent of the sign of the applied bias voltage, the tail-to-tail domain wall consistently moves towards its nearest head-to-head counterpart. This also applies for the head-to-head walls as shown by the PFM map in Fig.~\ref{fig:figure_5}(c). The same effect is observed when the tip is placed between the domain walls (Supplementary Fig. S6). 

One possible driving mechanism for this unusual switching behavior is the electric field acting on the secondary strain-capable polar $\Gamma_5$ mode, which interacts with the domain walls due to their ferroelastic nature. We consistently find that neighboring head-to-head and tail-to-tail walls approach each other, in agreement with their stress-promoted annihilation observed in optical experiments on much larger length scales~\cite{Voronkova2000}.

Most importantly for this study, Figure~\ref{fig:figure_5}(a,b) shows that tail-to-tail and head-to-head walls are spatially mobile. EFM data gained at the same position (Fig.~\ref{fig:figure_5}(d)) reveals that the electrostatic signature moves along with the walls, corroborating that it is an intrinsic feature. Similar to the EFM scan in Fig.~\ref{fig:figure_3}(d), the head-to-head domain wall exhibits a pronounced variation in surface potential, whereas only a very weak EFM signal is measured at the new position of the tail-to-tail wall (dark spots in the EFM map mark the positions where the biased tip was placed). As shown in Supplementary Fig. S6(c,d), head-to-tail domain walls are in general immobile (when no head-to-head and tail-to-tail domain walls are nearby).

In striking contrast to the straight domain walls in the unpoled state (Fig.~\ref{fig:figure_3}(a)), the careful evaluation of the VPFM data (Supplementary Note 2) in Fig.~\ref{fig:figure_5}(c) reveals that the electromechanical response is no longer constant along the wall after poling. As shown in Fig.~\ref{fig:figure_5}(e,f), in the electric-field-induced curved domain wall segment, the piezoresponse (VPFM signal) continuously varies together with the charge state of the domain wall ($\propto 2P_a \cos{\theta}$). This behavior is remarkable as it shows that the piezoresponse of the charged domain walls is highly tunable. A correlation between charge state and piezoresponse was previously reported for domain walls in improper ferroelectric HoMnO$_3$\cite{Lochocki2011}. In the latter case, however, the correlation was studied only for as-grown domain walls without showing the possibility to control the local piezoreponse by changing the domain wall orientation. 

Our analysis of the local electromechanical properties demonstrates that the piezoresponse of individual domain walls can be tuned, going from discrete softness states~\cite{Stefani2020} to a continuous set of states with gradually varying softness with the domain wall orientation as control parameters.

\subsection*{Outlook}

 The results establish domain walls in noncollinear antiferroelectrics as multifunctional 2D systems with unusual physical properties. Enabled by the symmetry-lowering caused by the canting of electric dipoles, these 2D systems unify properties that are usually specific to either ferroelectric or antiferroelectric materials, classifying them as hybrid antiferroelectric-ferroelectric (and ferroelastic) domain walls. The possibility to combine functional properties that otherwise arise in different types of materials --- together with the demonstrated tunability --- is of interest for the field of domain wall engineering and the design of advanced multifunctional domain-wall-based devices. Importantly, the concept of enhancing the physical responses of antiferroelectrics by introducing noncollinearity is not restricted to the case of K$_3$[Nb$_3$O$_6|$(BO$_3$)$_2$], and we expect similar effects to arise in a much larger class of materials. Another promising candidate system is the improper ferroelectric Gd$_2$(MoO$_4$)$_3$, where most of the electric dipole moments in the unit cell cancel out and a spontaneous polarization only appears due to a small canting \cite{Jeitschko1972}. Such noncollinear order is symmetry-allowed in all non-centrosymmetric systems with antipolar order in the direction perpendicular to the polar axis, facilitating, e.g., hybrid antiferroelectric-ferroelectric behavior as demonstrated in this work. Notwithstanding material-specific details concerning the coupling between polar and antipolar orders, this relationship provides us with a universal guideline where to search for noncollinear antiferroelectricity and noncollinearity-driven nanoscale phenomena. Noncollinearity may further be leveraged to enhance the functionality of antiferroelectric heterostructures and multilayers, giving a new dimension to the research on antiferroelectrics and anti-ferroic oxides in general.

\bibliography{sn-bibliography}


\begin{thebibliography}{46}
\ifx \bisbn   \undefined \def \bisbn  #1{ISBN #1}\fi
\ifx \binits  \undefined \def \binits#1{#1}\fi
\ifx \bauthor  \undefined \def \bauthor#1{#1}\fi
\ifx \batitle  \undefined \def \batitle#1{#1}\fi
\ifx \bjtitle  \undefined \def \bjtitle#1{#1}\fi
\ifx \bvolume  \undefined \def \bvolume#1{\textbf{#1}}\fi
\ifx \byear  \undefined \def \byear#1{#1}\fi
\ifx \bissue  \undefined \def \bissue#1{#1}\fi
\ifx \bfpage  \undefined \def \bfpage#1{#1}\fi
\ifx \blpage  \undefined \def \blpage #1{#1}\fi
\ifx \burl  \undefined \def \burl#1{\textsf{#1}}\fi
\ifx \doiurl  \undefined \def \doiurl#1{\url{https://doi.org/#1}}\fi
\ifx \betal  \undefined \def \betal{\textit{et al.}}\fi
\ifx \binstitute  \undefined \def \binstitute#1{#1}\fi
\ifx \binstitutionaled  \undefined \def \binstitutionaled#1{#1}\fi
\ifx \bctitle  \undefined \def \bctitle#1{#1}\fi
\ifx \beditor  \undefined \def \beditor#1{#1}\fi
\ifx \bpublisher  \undefined \def \bpublisher#1{#1}\fi
\ifx \bbtitle  \undefined \def \bbtitle#1{#1}\fi
\ifx \bedition  \undefined \def \bedition#1{#1}\fi
\ifx \bseriesno  \undefined \def \bseriesno#1{#1}\fi
\ifx \blocation  \undefined \def \blocation#1{#1}\fi
\ifx \bsertitle  \undefined \def \bsertitle#1{#1}\fi
\ifx \bsnm \undefined \def \bsnm#1{#1}\fi
\ifx \bsuffix \undefined \def \bsuffix#1{#1}\fi
\ifx \bparticle \undefined \def \bparticle#1{#1}\fi
\ifx \barticle \undefined \def \barticle#1{#1}\fi
\bibcommenthead
\ifx \bconfdate \undefined \def \bconfdate #1{#1}\fi
\ifx \botherref \undefined \def \botherref #1{#1}\fi
\ifx \url \undefined \def \url#1{\textsf{#1}}\fi
\ifx \bchapter \undefined \def \bchapter#1{#1}\fi
\ifx \bbook \undefined \def \bbook#1{#1}\fi
\ifx \bcomment \undefined \def \bcomment#1{#1}\fi
\ifx \oauthor \undefined \def \oauthor#1{#1}\fi
\ifx \citeauthoryear \undefined \def \citeauthoryear#1{#1}\fi
\ifx \endbibitem  \undefined \def \endbibitem {}\fi
\ifx \bconflocation  \undefined \def \bconflocation#1{#1}\fi
\ifx \arxivurl  \undefined \def \arxivurl#1{\textsf{#1}}\fi
\csname PreBibitemsHook\endcsname

%%% 1
\bibitem[\protect\citeauthoryear{N{\'e}el}{1972}]{Neel1972}
\begin{botherref}
\oauthor{\bsnm{N{\'e}el}, \binits{L.}}:
Magnetism and the local molecular field.
Nobel Lectures Physics 1963-1970
(1972)
\end{botherref}
\endbibitem

%%% 2
\bibitem[\protect\citeauthoryear{Kittel}{1951}]{Kittel1951}
\begin{barticle}
\bauthor{\bsnm{Kittel}, \binits{C.}}:
\batitle{Theory of antiferroelectric crystals}.
\bjtitle{Physical Review}
\bvolume{82}(\bissue{5}),
\bfpage{729}
(\byear{1951})
\end{barticle}
\endbibitem

%%% 3
\bibitem[\protect\citeauthoryear{Jungwirth et~al.}{2016}]{Jungwirth2016}
\begin{barticle}
\bauthor{\bsnm{Jungwirth}, \binits{T.}},
\bauthor{\bsnm{Marti}, \binits{X.}},
\bauthor{\bsnm{Wadley}, \binits{P.}},
\bauthor{\bsnm{Wunderlich}, \binits{J.}}:
\batitle{Antiferromagnetic spintronics}.
\bjtitle{Nature Nanotechnology}
\bvolume{11}(\bissue{3}),
\bfpage{231}--\blpage{241}
(\byear{2016})
\end{barticle}
\endbibitem

%%% 4
\bibitem[\protect\citeauthoryear{Han et~al.}{2023}]{Han2023}
\begin{barticle}
\bauthor{\bsnm{Han}, \binits{J.}},
\bauthor{\bsnm{Cheng}, \binits{R.}},
\bauthor{\bsnm{Liu}, \binits{L.}},
\bauthor{\bsnm{Ohno}, \binits{H.}},
\bauthor{\bsnm{Fukami}, \binits{S.}}:
\batitle{Coherent antiferromagnetic spintronics}.
\bjtitle{Nature Materials}
\bvolume{22}(\bissue{6}),
\bfpage{684}--\blpage{695}
(\byear{2023})
\end{barticle}
\endbibitem

%%% 5
\bibitem[\protect\citeauthoryear{Liu et~al.}{2018}]{Liu2018}
\begin{barticle}
\bauthor{\bsnm{Liu}, \binits{Z.}},
\bauthor{\bsnm{Lu}, \binits{T.}},
\bauthor{\bsnm{Ye}, \binits{J.}},
\bauthor{\bsnm{Wang}, \binits{G.}},
\bauthor{\bsnm{Dong}, \binits{X.}},
\bauthor{\bsnm{Withers}, \binits{R.}},
\bauthor{\bsnm{Liu}, \binits{Y.}}:
\batitle{Antiferroelectrics for energy storage applications: a review}.
\bjtitle{Advanced Materials Technologies}
\bvolume{3}(\bissue{9}),
\bfpage{1800111}
(\byear{2018})
\end{barticle}
\endbibitem

%%% 6
\bibitem[\protect\citeauthoryear{Randall et~al.}{2021}]{Randall2021}
\begin{barticle}
\bauthor{\bsnm{Randall}, \binits{C.A.}},
\bauthor{\bsnm{Fan}, \binits{Z.}},
\bauthor{\bsnm{Reaney}, \binits{I.}},
\bauthor{\bsnm{Chen}, \binits{L.-Q.}},
\bauthor{\bsnm{Trolier-McKinstry}, \binits{S.}}:
\batitle{Antiferroelectrics: History, fundamentals, crystal chemistry, crystal structures, size effects, and applications}.
\bjtitle{Journal of the American Ceramic Society}
\bvolume{104}(\bissue{8}),
\bfpage{3775}--\blpage{3810}
(\byear{2021})
\end{barticle}
\endbibitem

%%% 7
\bibitem[\protect\citeauthoryear{Blinc and {\v{Z}}ek{\v{s}}}{1974}]{Blinc1974}
\begin{bbook}
\bauthor{\bsnm{Blinc}, \binits{R.}},
\bauthor{\bsnm{{\v{Z}}ek{\v{s}}}, \binits{B.}}:
\bbtitle{Soft Modes in Ferroelectrics and Antiferroelectrics}.
\bsertitle{Science Education and Future Human Needs Series}.
\bpublisher{North-Holland Publishing Company},
\blocation{Amsterdam, Netherlands}
(\byear{1974})
\end{bbook}
\endbibitem

%%% 8
\bibitem[\protect\citeauthoryear{Zheludev}{1971}]{Zheludev1971}
\begin{bchapter}
\bauthor{\bsnm{Zheludev}, \binits{I.}}:
\bctitle{Ferroelectricity and symmetry}.
In: \bbtitle{Solid State Physics}
vol. \bseriesno{26},
pp. \bfpage{429}--\blpage{464}.
\bpublisher{Elsevier},
\blocation{Amsterdam, Netherlands}
(\byear{1971})
\end{bchapter}
\endbibitem

%%% 9
\bibitem[\protect\citeauthoryear{Scott}{1974}]{Scott1974}
\begin{barticle}
\bauthor{\bsnm{Scott}, \binits{J.}}:
\batitle{Soft-mode spectroscopy: Experimental studies of structural phase transitions}.
\bjtitle{Reviews of Modern Physics}
\bvolume{46}(\bissue{1}),
\bfpage{83}
(\byear{1974})
\end{barticle}
\endbibitem

%%% 10
\bibitem[\protect\citeauthoryear{Jeitschko}{1972}]{Jeitschko1972}
\begin{barticle}
\bauthor{\bsnm{Jeitschko}, \binits{W.}}:
\batitle{A comprehensive x-ray study of the ferroelectric--ferroelastic and paraelectric--paraelastic phases of {Gd}$_2$({MoO}$_4$)$_3$}.
\bjtitle{Acta Crystallographica Section B: Structural Crystallography and Crystal Chemistry}
\bvolume{28}(\bissue{1}),
\bfpage{60}--\blpage{76}
(\byear{1972})
\end{barticle}
\endbibitem

%%% 11
\bibitem[\protect\citeauthoryear{Khalyavin et~al.}{2020}]{Khalyavin2020}
\begin{barticle}
\bauthor{\bsnm{Khalyavin}, \binits{D.D.}},
\bauthor{\bsnm{Johnson}, \binits{R.D.}},
\bauthor{\bsnm{Orlandi}, \binits{F.}},
\bauthor{\bsnm{Radaelli}, \binits{P.G.}},
\bauthor{\bsnm{Manuel}, \binits{P.}},
\bauthor{\bsnm{Belik}, \binits{A.A.}}:
\batitle{Emergent helical texture of electric dipoles}.
\bjtitle{Science}
\bvolume{369}(\bissue{6504}),
\bfpage{680}--\blpage{684}
(\byear{2020})
\end{barticle}
\endbibitem

%%% 12
\bibitem[\protect\citeauthoryear{Gao et~al.}{2022}]{Gao2022}
\begin{barticle}
\bauthor{\bsnm{Gao}, \binits{B.}},
\bauthor{\bsnm{Liu}, \binits{H.}},
\bauthor{\bsnm{Zhou}, \binits{Z.}},
\bauthor{\bsnm{Xu}, \binits{K.}},
\bauthor{\bsnm{Qi}, \binits{H.}},
\bauthor{\bsnm{Deng}, \binits{S.}},
\bauthor{\bsnm{Ren}, \binits{Y.}},
\bauthor{\bsnm{Sun}, \binits{J.}},
\bauthor{\bsnm{Huang}, \binits{H.}},
\bauthor{\bsnm{Chen}, \binits{J.}}:
\batitle{An intriguing canting dipole configuration and its evolution under an electric field in {La}-doped {Pb}({Zr}, {Sn}, {Ti}){O}$_3$ perovskites}.
\bjtitle{Microstructures}
\bvolume{2}(\bissue{2}),
\bfpage{2022010}
(\byear{2022})
\end{barticle}
\endbibitem

%%% 13
\bibitem[\protect\citeauthoryear{Wang et~al.}{2024}]{Wang2024}
\begin{barticle}
\bauthor{\bsnm{Wang}, \binits{N.}},
\bauthor{\bsnm{Shen}, \binits{Z.}},
\bauthor{\bsnm{Luo}, \binits{W.}},
\bauthor{\bsnm{Li}, \binits{H.-K.}},
\bauthor{\bsnm{Xu}, \binits{Z.-J.}},
\bauthor{\bsnm{Shi}, \binits{C.}},
\bauthor{\bsnm{Ye}, \binits{H.-Y.}},
\bauthor{\bsnm{Dong}, \binits{S.}},
\bauthor{\bsnm{Miao}, \binits{L.-P.}}:
\batitle{Noncollinear ferroelectric and screw-type antiferroelectric phases in a metal-free hybrid molecular crystal}.
\bjtitle{Nature Communications}
\bvolume{15}(\bissue{1}),
\bfpage{10262}
(\byear{2024})
\end{barticle}
\endbibitem

%%% 14
\bibitem[\protect\citeauthoryear{Zhao et~al.}{2021}]{Zhao2021}
\begin{barticle}
\bauthor{\bsnm{Zhao}, \binits{H.J.}},
\bauthor{\bsnm{Chen}, \binits{P.}},
\bauthor{\bsnm{Prosandeev}, \binits{S.}},
\bauthor{\bsnm{Artyukhin}, \binits{S.}},
\bauthor{\bsnm{Bellaiche}, \binits{L.}}:
\batitle{Dzyaloshinskii--moriya-like interaction in ferroelectrics and antiferroelectrics}.
\bjtitle{Nature Materials}
\bvolume{20}(\bissue{3}),
\bfpage{341}--\blpage{345}
(\byear{2021})
\end{barticle}
\endbibitem

%%% 15
\bibitem[\protect\citeauthoryear{Becker et~al.}{1996}]{Becker1996}
\begin{barticle}
\bauthor{\bsnm{Becker}, \binits{P.}},
\bauthor{\bsnm{Held}, \binits{P.}},
\bauthor{\bsnm{Bohat{\'y}}, \binits{L.}}:
\batitle{Crystal growth of ferroelectric and ferroelastic {K}$_3$[{Nb}$_3${O}$_6$({BO}$_3$)$_2$] and crystal structure of the room temperature modification}.
\bjtitle{Zeitschrift f{\"u}r Kristallographie-Crystalline Materials}
\bvolume{211}(\bissue{7}),
\bfpage{449}--\blpage{452}
(\byear{1996})
\end{barticle}
\endbibitem

%%% 16
\bibitem[\protect\citeauthoryear{Becker et~al.}{1997}]{Becker1997}
\begin{barticle}
\bauthor{\bsnm{Becker}, \binits{P.}},
\bauthor{\bsnm{Bohat{\'y}}, \binits{L.}},
\bauthor{\bsnm{Schneider}, \binits{J.}}:
\batitle{Rietveld refinement of the structure of high-temperature {K}$_3$[{Nb}$_3${O}$_6$({BO}$_3$)$_2$]}.
\bjtitle{Kristallografiya}
\bvolume{42}(\bissue{2}),
\bfpage{250}--\blpage{254}
(\byear{1997})
\end{barticle}
\endbibitem

%%% 17
\bibitem[\protect\citeauthoryear{Aizu}{1970}]{Aizu1970}
\begin{barticle}
\bauthor{\bsnm{Aizu}, \binits{K.}}:
\batitle{Determination of the state parameters and formulation of spontaneous strain for ferroelastics}.
\bjtitle{Journal of the Physical Society of Japan}
\bvolume{28}(\bissue{3}),
\bfpage{706}--\blpage{716}
(\byear{1970})
\end{barticle}
\endbibitem

%%% 18
\bibitem[\protect\citeauthoryear{Erb and Hlinka}{2020}]{Erb2020}
\begin{barticle}
\bauthor{\bsnm{Erb}, \binits{K.}},
\bauthor{\bsnm{Hlinka}, \binits{J.}}:
\batitle{Vector, bidirector, and {B}loch skyrmion phases induced by structural crystallographic symmetry breaking}.
\bjtitle{Physical Review B}
\bvolume{102}(\bissue{2}),
\bfpage{024110}
(\byear{2020})
\end{barticle}
\endbibitem

%%% 19
\bibitem[\protect\citeauthoryear{Authier}{2014}]{Authier2014}
\begin{bbook}
\beditor{\bsnm{Authier}, \binits{A.}} (ed.):
\bbtitle{International Tables for Crystallography, Volume D: Physical Properties of Crystals}
vol. \bseriesno{4}.
\bpublisher{John Wiley \& Sons},
\blocation{Hoboken, NJ, USA}
(\byear{2014})
\end{bbook}
\endbibitem

%%% 20
\bibitem[\protect\citeauthoryear{Sapriel}{1975}]{Sapriel1975}
\begin{barticle}
\bauthor{\bsnm{Sapriel}, \binits{J.}}:
\batitle{Domain-wall orientations in ferroelastics}.
\bjtitle{Physical Review B}
\bvolume{12}(\bissue{11}),
\bfpage{5128}
(\byear{1975})
\end{barticle}
\endbibitem

%%% 21
\bibitem[\protect\citeauthoryear{Erhart}{2004}]{Erhart2004}
\begin{barticle}
\bauthor{\bsnm{Erhart}, \binits{J.}}:
\batitle{Domain wall orientations in ferroelastics and ferroelectrics}.
\bjtitle{Phase Transitions}
\bvolume{77}(\bissue{12}),
\bfpage{989}--\blpage{1074}
(\byear{2004})
\end{barticle}
\endbibitem

%%% 22
\bibitem[\protect\citeauthoryear{Kaminskii et~al.}{2004}]{Kaminskii2004}
\begin{barticle}
\bauthor{\bsnm{Kaminskii}, \binits{A.}},
\bauthor{\bsnm{Becker}, \binits{P.}},
\bauthor{\bsnm{Bohat{\'y}}, \binits{L.}},
\bauthor{\bsnm{Eichler}, \binits{H.}},
\bauthor{\bsnm{Penin}, \binits{A.}},
\bauthor{\bsnm{Ueda}, \binits{K.}},
\bauthor{\bsnm{Hanuza}, \binits{J.}},
\bauthor{\bsnm{Takaichi}, \binits{K.}},
\bauthor{\bsnm{Rhee}, \binits{H.}}:
\batitle{Room-temperature high-order stokes and anti-stokes generation in orthorhombic ferroelectric-ferroelastic {K}$_3${Nb}$_3${O}$_6$({BO}$_3$)$_2$ crystal}.
\bjtitle{physica status solidi (a)}
\bvolume{201}(\bissue{9}),
\bfpage{2154}--\blpage{2169}
(\byear{2004})
\end{barticle}
\endbibitem

%%% 23
\bibitem[\protect\citeauthoryear{Campbell et~al.}{2006}]{Campbell2006}
\begin{barticle}
\bauthor{\bsnm{Campbell}, \binits{B.J.}},
\bauthor{\bsnm{Stokes}, \binits{H.T.}},
\bauthor{\bsnm{Tanner}, \binits{D.E.}},
\bauthor{\bsnm{Hatch}, \binits{D.M.}}:
\batitle{Isodisplace: a web-based tool for exploring structural distortions}.
\bjtitle{Journal of Applied Crystallography}
\bvolume{39}(\bissue{4}),
\bfpage{607}--\blpage{614}
(\byear{2006})
\end{barticle}
\endbibitem

%%% 24
\bibitem[\protect\citeauthoryear{Hatch and Stokes}{2003}]{Hatch2003}
\begin{barticle}
\bauthor{\bsnm{Hatch}, \binits{D.M.}},
\bauthor{\bsnm{Stokes}, \binits{H.T.}}:
\batitle{Invariants: program for obtaining a list of invariant polynomials of the order-parameter components associated with irreducible representations of a space group}.
\bjtitle{Journal of Applied Crystallography}
\bvolume{36}(\bissue{3}),
\bfpage{951}--\blpage{952}
(\byear{2003})
\end{barticle}
\endbibitem

%%% 25
\bibitem[\protect\citeauthoryear{Merz}{1949}]{Merz1949}
\begin{barticle}
\bauthor{\bsnm{Merz}, \binits{W.J.}}:
\batitle{The electric and optical behavior of {Ba}{Ti}{O}$_3$ single-domain crystals}.
\bjtitle{Physical Review}
\bvolume{76}(\bissue{8}),
\bfpage{1221}
(\byear{1949})
\end{barticle}
\endbibitem

%%% 26
\bibitem[\protect\citeauthoryear{Schmid et~al.}{1980}]{Schmid1980}
\begin{barticle}
\bauthor{\bsnm{Schmid}, \binits{H.}},
\bauthor{\bsnm{Genequand}, \binits{P.}},
\bauthor{\bsnm{Pouilly}, \binits{G.}},
\bauthor{\bsnm{Chan}, \binits{P.}}:
\batitle{Pyroelectricity of {Fe}-{I} and {Cu}-{Cl} boracite}.
\bjtitle{Ferroelectrics}
\bvolume{25}(\bissue{1}),
\bfpage{539}--\blpage{542}
(\byear{1980})
\end{barticle}
\endbibitem

%%% 27
\bibitem[\protect\citeauthoryear{Cummins}{1970}]{Cummins1970}
\begin{barticle}
\bauthor{\bsnm{Cummins}, \binits{S.}}:
\batitle{Electrical, optical, and mechanical behavior of ferroelectric {Gd}$_2$({MoO}$_4$)$_3$}.
\bjtitle{Ferroelectrics}
\bvolume{1}(\bissue{1}),
\bfpage{11}--\blpage{17}
(\byear{1970})
\end{barticle}
\endbibitem

%%% 28
\bibitem[\protect\citeauthoryear{Shan et~al.}{2020}]{Shan2020}
\begin{barticle}
\bauthor{\bsnm{Shan}, \binits{P.}},
\bauthor{\bsnm{Xiong}, \binits{J.}},
\bauthor{\bsnm{Wang}, \binits{Z.}},
\bauthor{\bsnm{He}, \binits{C.}},
\bauthor{\bsnm{Yang}, \binits{X.}},
\bauthor{\bsnm{Su}, \binits{R.}},
\bauthor{\bsnm{Long}, \binits{X.}}:
\batitle{Lead-free polar borate crystal {K}$_3${Nb}$_3${B}$_2${O}$_{12}$: a novel antiferroelectric structure type}.
\bjtitle{Journal of Materials Chemistry C}
\bvolume{8}(\bissue{20}),
\bfpage{6654}--\blpage{6658}
(\byear{2020})
\end{barticle}
\endbibitem

%%% 29
\bibitem[\protect\citeauthoryear{Park et~al.}{1997}]{Park1997}
\begin{barticle}
\bauthor{\bsnm{Park}, \binits{S.-E.}},
\bauthor{\bsnm{Pan}, \binits{M.-J.}},
\bauthor{\bsnm{Markowski}, \binits{K.}},
\bauthor{\bsnm{Yoshikawa}, \binits{S.}},
\bauthor{\bsnm{Cross}, \binits{L.E.}}:
\batitle{Electric field induced phase transition of antiferroelectric lead lanthanum zirconate titanate stannate ceramics}.
\bjtitle{Journal of Applied Physics}
\bvolume{82}(\bissue{4}),
\bfpage{1798}--\blpage{1803}
(\byear{1997})
\end{barticle}
\endbibitem

%%% 30
\bibitem[\protect\citeauthoryear{Fesenko et~al.}{1978}]{Fesenko1978}
\begin{barticle}
\bauthor{\bsnm{Fesenko}, \binits{O.}},
\bauthor{\bsnm{Kolesova}, \binits{R.}},
\bauthor{\bsnm{Sindeyev}, \binits{Y.G.}}:
\batitle{The structural phase transitions in lead zirconate in super-high electric fields}.
\bjtitle{Ferroelectrics}
\bvolume{20}(\bissue{1}),
\bfpage{177}--\blpage{178}
(\byear{1978})
\end{barticle}
\endbibitem

%%% 31
\bibitem[\protect\citeauthoryear{Shirane et~al.}{1951}]{Shirane1951}
\begin{barticle}
\bauthor{\bsnm{Shirane}, \binits{G.}},
\bauthor{\bsnm{Sawaguchi}, \binits{E.}},
\bauthor{\bsnm{Takagi}, \binits{Y.}}:
\batitle{Dielectric properties of lead zirconate}.
\bjtitle{Physical Review}
\bvolume{84}(\bissue{3}),
\bfpage{476}
(\byear{1951})
\end{barticle}
\endbibitem

%%% 32
\bibitem[\protect\citeauthoryear{Teslic and Egami}{1998}]{Teslic1998}
\begin{barticle}
\bauthor{\bsnm{Teslic}, \binits{S.}},
\bauthor{\bsnm{Egami}, \binits{T.}}:
\batitle{Atomic structure of {PbZrO}$_3$ determined by pulsed neutron diffraction}.
\bjtitle{Acta Crystallographica Section B: Structural Science}
\bvolume{54}(\bissue{6}),
\bfpage{750}--\blpage{765}
(\byear{1998})
\end{barticle}
\endbibitem

%%% 33
\bibitem[\protect\citeauthoryear{Salje}{1991}]{Salje1991}
\begin{bbook}
\bauthor{\bsnm{Salje}, \binits{E.K.}}:
\bbtitle{Phase Transitions in Ferroelastic and Co-elastic Crystals}.
\bsertitle{Cambridge Topics in Mineral Physics and Chemistry}.
\bpublisher{Cambridge University Press},
\blocation{Cambridge}
(\byear{1991})
\end{bbook}
\endbibitem

%%% 34
\bibitem[\protect\citeauthoryear{Sui and Huber}{2020}]{Sui2020}
\begin{barticle}
\bauthor{\bsnm{Sui}, \binits{D.}},
\bauthor{\bsnm{Huber}, \binits{J.E.}}:
\batitle{Modelling and interaction of needle domains in barium titanate single crystals}.
\bjtitle{European Journal of Mechanics-A/Solids}
\bvolume{80},
\bfpage{103919}
(\byear{2020})
\end{barticle}
\endbibitem

%%% 35
\bibitem[\protect\citeauthoryear{McQuaid et~al.}{2017}]{Mcquaid2017}
\begin{barticle}
\bauthor{\bsnm{McQuaid}, \binits{R.G.}},
\bauthor{\bsnm{Campbell}, \binits{M.P.}},
\bauthor{\bsnm{Whatmore}, \binits{R.W.}},
\bauthor{\bsnm{Kumar}, \binits{A.}},
\bauthor{\bsnm{Gregg}, \binits{J.M.}}:
\batitle{Injection and controlled motion of conducting domain walls in improper ferroelectric {Cu}-{Cl} boracite}.
\bjtitle{Nature Communications}
\bvolume{8}(\bissue{1}),
\bfpage{15105}
(\byear{2017})
\end{barticle}
\endbibitem

%%% 36
\bibitem[\protect\citeauthoryear{Lu et~al.}{2023}]{Lu2023}
\begin{botherref}
\oauthor{\bsnm{Lu}, \binits{H.}},
\oauthor{\bsnm{Tan}, \binits{Y.}},
\oauthor{\bsnm{Richarz}, \binits{L.}},
\oauthor{\bsnm{He}, \binits{J.}},
\oauthor{\bsnm{Wang}, \binits{B.}},
\oauthor{\bsnm{Meier}, \binits{D.}},
\oauthor{\bsnm{Chen}, \binits{L.-Q.}},
\oauthor{\bsnm{Gruverman}, \binits{A.}}:
Electromechanics of domain walls in uniaxial ferroelectrics.
Advanced Functional Materials,
2213684
(2023)
\end{botherref}
\endbibitem

%%% 37
\bibitem[\protect\citeauthoryear{Meier et~al.}{2012}]{Meier2012}
\begin{barticle}
\bauthor{\bsnm{Meier}, \binits{D.}},
\bauthor{\bsnm{Seidel}, \binits{J.}},
\bauthor{\bsnm{Cano}, \binits{A.}},
\bauthor{\bsnm{Delaney}, \binits{K.}},
\bauthor{\bsnm{Kumagai}, \binits{Y.}},
\bauthor{\bsnm{Mostovoy}, \binits{M.}},
\bauthor{\bsnm{Spaldin}, \binits{N.A.}},
\bauthor{\bsnm{Ramesh}, \binits{R.}},
\bauthor{\bsnm{Fiebig}, \binits{M.}}:
\batitle{Anisotropic conductance at improper ferroelectric domain walls}.
\bjtitle{Nature materials}
\bvolume{11}(\bissue{4}),
\bfpage{284}--\blpage{288}
(\byear{2012})
\end{barticle}
\endbibitem

%%% 38
\bibitem[\protect\citeauthoryear{Eliseev et~al.}{2011}]{Eliseev2011}
\begin{barticle}
\bauthor{\bsnm{Eliseev}, \binits{E.}},
\bauthor{\bsnm{Morozovska}, \binits{A.}},
\bauthor{\bsnm{Svechnikov}, \binits{G.}},
\bauthor{\bsnm{Gopalan}, \binits{V.}},
\bauthor{\bsnm{Shur}, \binits{V.Y.}}:
\batitle{Static conductivity of charged domain walls in uniaxial ferroelectric semiconductors}.
\bjtitle{Physical Review B—Condensed Matter and Materials Physics}
\bvolume{83}(\bissue{23}),
\bfpage{235313}
(\byear{2011})
\end{barticle}
\endbibitem

%%% 39
\bibitem[\protect\citeauthoryear{Gureev et~al.}{2011}]{Gureev2011}
\begin{barticle}
\bauthor{\bsnm{Gureev}, \binits{M.Y.}},
\bauthor{\bsnm{Tagantsev}, \binits{A.K.}},
\bauthor{\bsnm{Setter}, \binits{N.}}:
\batitle{Head-to-head and tail-to-tail 180{$^\circ$} domain walls in an isolated ferroelectric}.
\bjtitle{Physical Review B—Condensed Matter and Materials Physics}
\bvolume{83}(\bissue{18}),
\bfpage{184104}
(\byear{2011})
\end{barticle}
\endbibitem

%%% 40
\bibitem[\protect\citeauthoryear{Tikhonov et~al.}{2022}]{Tikhonov2022}
\begin{barticle}
\bauthor{\bsnm{Tikhonov}, \binits{Y.}},
\bauthor{\bsnm{Maguire}, \binits{J.R.}},
\bauthor{\bsnm{McCluskey}, \binits{C.J.}},
\bauthor{\bsnm{McConville}, \binits{J.P.}},
\bauthor{\bsnm{Kumar}, \binits{A.}},
\bauthor{\bsnm{Lu}, \binits{H.}},
\bauthor{\bsnm{Meier}, \binits{D.}},
\bauthor{\bsnm{Razumnaya}, \binits{A.}},
\bauthor{\bsnm{Gregg}, \binits{J.M.}},
\bauthor{\bsnm{Gruverman}, \binits{A.}}, \betal:
\batitle{Polarization topology at the nominally charged domain walls in uniaxial ferroelectrics}.
\bjtitle{Advanced Materials}
\bvolume{34}(\bissue{45}),
\bfpage{2203028}
(\byear{2022})
\end{barticle}
\endbibitem

%%% 41
\bibitem[\protect\citeauthoryear{Voronkova et~al.}{2000}]{Voronkova2000}
\begin{barticle}
\bauthor{\bsnm{Voronkova}, \binits{V.}},
\bauthor{\bsnm{Kharitonova}, \binits{E.}},
\bauthor{\bsnm{Yanovskii}, \binits{V.}},
\bauthor{\bsnm{Stefanovich}, \binits{S.Y.}},
\bauthor{\bsnm{Mosunov}, \binits{A.}},
\bauthor{\bsnm{Sorokina}, \binits{N.}}:
\batitle{Growth, structure, and properties of ferroelectric—ferroelastic—superionic {K}$_3${Nb}$_3${B}$_2${O}$_{12}$ and {K}$_{3-x}${Na}$_x${Nb}$_3${B}$_2${O}$_{12}$ crystals}.
\bjtitle{Crystallography Reports}
\bvolume{45},
\bfpage{816}--\blpage{820}
(\byear{2000})
\end{barticle}
\endbibitem

%%% 42
\bibitem[\protect\citeauthoryear{Lochocki et~al.}{2011}]{Lochocki2011}
\begin{barticle}
\bauthor{\bsnm{Lochocki}, \binits{E.B.}},
\bauthor{\bsnm{Park}, \binits{S.}},
\bauthor{\bsnm{Lee}, \binits{N.}},
\bauthor{\bsnm{Cheong}, \binits{S.-W.}},
\bauthor{\bsnm{Wu}, \binits{W.}}:
\batitle{Piezoresponse force microscopy of domains and walls in multiferroic {Ho}{Mn}{O}$_3$}.
\bjtitle{Applied Physics Letters}
\bvolume{99}(\bissue{23}),
\bfpage{232901}
(\byear{2011})
\end{barticle}
\endbibitem

%%% 43
\bibitem[\protect\citeauthoryear{Stefani et~al.}{2020}]{Stefani2020}
\begin{barticle}
\bauthor{\bsnm{Stefani}, \binits{C.}},
\bauthor{\bsnm{Ponet}, \binits{L.}},
\bauthor{\bsnm{Shapovalov}, \binits{K.}},
\bauthor{\bsnm{Chen}, \binits{P.}},
\bauthor{\bsnm{Langenberg}, \binits{E.}},
\bauthor{\bsnm{Schlom}, \binits{D.G.}},
\bauthor{\bsnm{Artyukhin}, \binits{S.}},
\bauthor{\bsnm{Stengel}, \binits{M.}},
\bauthor{\bsnm{Domingo}, \binits{N.}},
\bauthor{\bsnm{Catalan}, \binits{G.}}:
\batitle{Mechanical softness of ferroelectric 180° domain walls}.
\bjtitle{Physical Review X}
\bvolume{10}(\bissue{4}),
\bfpage{041001}
(\byear{2020})
\end{barticle}
\endbibitem

%%% 44
\bibitem[\protect\citeauthoryear{Soergel}{2011}]{Soergel2011}
\begin{barticle}
\bauthor{\bsnm{Soergel}, \binits{E.}}:
\batitle{Piezoresponse force microscopy {(PFM)}}.
\bjtitle{Journal of Physics D: Applied Physics}
\bvolume{44}(\bissue{46}),
\bfpage{464003}
(\byear{2011})
\end{barticle}
\endbibitem

%%% 45
\bibitem[\protect\citeauthoryear{Eberg et~al.}{2008}]{Eberg2008}
\begin{barticle}
\bauthor{\bsnm{Eberg}, \binits{E.}},
\bauthor{\bsnm{Monsen}, \binits{{\AA}.F.}},
\bauthor{\bsnm{Tybell}, \binits{T.}},
\bauthor{\bsnm{Helvoort}, \binits{A.T.}},
\bauthor{\bsnm{Holmestad}, \binits{R.}}:
\batitle{Comparison of {TEM} specimen preparation of perovskite thin films by tripod polishing and conventional ion milling}.
\bjtitle{Journal of Electron Microscopy}
\bvolume{57}(\bissue{6}),
\bfpage{175}--\blpage{179}
(\byear{2008})
\end{barticle}
\endbibitem

%%% 46
\bibitem[\protect\citeauthoryear{Schaffer et~al.}{2012}]{Schaffer2012}
\begin{barticle}
\bauthor{\bsnm{Schaffer}, \binits{M.}},
\bauthor{\bsnm{Schaffer}, \binits{B.}},
\bauthor{\bsnm{Ramasse}, \binits{Q.}}:
\batitle{Sample preparation for atomic-resolution {STEM} at low voltages by {FIB}}.
\bjtitle{Ultramicroscopy}
\bvolume{114},
\bfpage{62}--\blpage{71}
(\byear{2012})
\end{barticle}
\endbibitem

\end{thebibliography}

\section{Methods}\label{sec11}

\textbf{Macroscopic electric-field switching}

$P(E)$ measurements were performed with a Keithley 6517B electrometer on a (001)-oriented platelet, coated with contacts of silver paint in a customized low-temperature setup based on a Displex closed cycle He-refrigerator. Due to experimental limitations, a maximum electric field of 15.9 kV/cm could be applied. The polarization was measured with two consecutive passes of the electric field (up-down-up-down), and all data shown are hysteresis loops from the second pass, after subtracting a small time-dependent drift of the measured charge.

\textbf{Scanning probe microscopy}

For all the SPM experiments, the samples were polished on a Logitech PM5 polishing machine with a PP5 jig, using Logitech 9 µm calcined aluminium oxide powder as lapping fluid, followed by SF1 polishing suspension. For the imaging of a face perpendicular to one of the a-axes (inset in Fig. \ref{fig:figure_2}(a)), the sample was cut using a Well 3500 diamond wire saw before polishing.

Lateral PFM was performed with NT-MDT NTEGRA Prima SPM (SFV102NTF head) and an external Stanford SR830 lock-in amplifier, using a diamond coated TipsNano DCP10 tip with a stiffness of about 15 nN/nm. 5-10 V AC-bias at 40.13 kHz was applied to the tip, well below the contact resonance, and the phase offset of the lock-in amplifier was calibrated relative to a poled LiNbO$_3$ sample \cite{Soergel2011}. The contact force during the scans was about 700 nN. The vector PFM procedure is described in Supplementary Note 2.

Vertical PFM, localized switching, and EFM were performed with an Oxford Instruments Cypher ES Environmental SPM, using a Ti/Ir coated Oxford Instruments ASYELEC.01-R2 tip with a stiffness of about 2.4 nN/nm. For the VPFM, 5-10 V AC-bias at contact resonance (about 360 kHz) was applied to the tip. The images were compared to LiNbO$_3$-calibrated off-resonance VPFM images, and the phase offset was adjusted accordingly. For the switching, different types of biases were applied to the tip while being in contact with the sample (more details in Supplementary Note 2). The contact force during scans and switching was about 100 nN. EFM was performed in dual-pass non-contact mode, with a lift height of 50 nm and tip DC-bias of 10 V (no AC-bias) applied during second pass.

Tomographic AFM (Figure \ref{fig:figure_2}(f)) was conducted on the (001) sample face using an Asylum research MFP-3D Infinity AFM system. The experiment was conducted in two sequential steps. In the first step, a B-doped diamond tip (Adama Technologies) with stiffness of about 350 nN/nm was used in the high deflection setpoint regime to remove the material layer by layer. The second step involved use of a Pt/Ir coated silicon tip (stiffness of about 2.8 nN/nm) to obtain good quality PFM images on the area of interest. These steps were repeated to obtain PFM images at different depths till a final depth of 4 µm was reached. During the tip milling step, the scan size was set to 40x40 µm, and PFM at an AC frequency of 4 MHz (near contact resonance of the Diamond tip) was performed simultaneously (albeit noisy) to have an overall idea of the domain wall propagation into the depth of the material. After every milling step, clear PFM images were obtained (Supplementary Note 2) through the Pt/Ir coated tips by applying an AC bias of 5-10 V at frequencies near the tip-surface contact resonance which was around 690 kHz for lateral and 320 kHz for vertical PFM.

\textbf{Transmission electron microscopy, specimen preparation and data collection}

Tripod wedge TEM specimens were made on an Allied Multiprep System as described in Ref. \cite{Eberg2008}. Diamond lapping films (DLF) from 15 down to 0.1 $\mu$m grain size were used with a wedge angle of $2^{\circ}$. Final polishing was done with 20 nm silica slurry. No further ion milling was applied. For correlated PLM-PFM-TEM (Supplementary Fig. S4) the specimen was regularly checked with optical microscopy in the final grinding steps until the region of interest was reached. DLFs were used for tripod polishing of TEM specimens. In addition, specimens were made by focus ion milling (FIB) using a Thermo Fisher Scientific G4 UX DualBeam FIB, as outlined in \cite{Schaffer2012}. In situ lift-out was done with backside milling and a final polishing voltage of 2 kV. (S)TEM was performed with a double spherical-aberration corrected cold FEG JEOL ARM 200FC, operated at 200 kV. High-resolution HAADF-STEM images were taken with a spatial resolution of 78 pm. HAADF-STEM images included in this work were acquired with a beam semiconvergence angle of 27 mrad, inner and outer semicollection angles of 43 and 170 mrad, and a beam current of 22 pA.

\bmhead{Supplementary information}

The article has two accompanying Supplementary Notes and nine Supplementary Figures.

%If your article has accompanying supplementary file/s please state so here. 

%Authors reporting data from electrophoretic gels and blots should supply the full unprocessed scans for key as part of their Supplementary information. This may be requested by the editorial team/s if it is missing.

%Please refer to Journal-level guidance for any specific requirements.

\bmhead{Acknowledgements}

The authors acknowledge fruitful discussions with Michael Carpenter, Jiří Hlinka, Pierre-Eymeric Janolin, Oskar Ryggetangen, Marthe Linnerud, Emil Christiansen and Thomas Tybell. I.N.U., K.A.H., and D.M. thank the Department of Materials Science and Engineering at NTNU for direct financial support and D.M. further thanks NTNU for support through the Onsager Fellowship Program and the outstanding Academic Fellow Program. U.L., J.H., I.N.U., and D.M. acknowledge funding from the European Research Council (ERC) under the European Union’s Horizon 2020 Research and Innovation Program (Grant Agreement No. 863691). The Research Council of Norway is acknowledged for the support to the Norwegian Micro- and Nano-Fabrication Facility, NorFab, project no. 245963 and Norwegian Centre for Transmission Electron Microscopy, NORTEM, Grant no. 197405. Work at LIST was financed by the Luxembourg National Research Fund (FNR) through Grant C21/MS/15799044/FERRODYNAMICS. A.K. and N.S. gratefully acknowledge support from the Department of Education and Learning NI through grant USI-211. Finite element modeling is also conducted through Project no 301020 financially supported by Norwegian Research Council. Computational resources were provided by UNINETT Sigma2 through Project NN9264K. M.Z.K and S.M.S. acknowledge support from the Research Council of Norway Project no 302506. J.H. and Ch.G. acknowledge funding from the Deutsche Forschungsgemeinschaft (DFG, German Research Foundation) through Project No. 277146847 – CRC 1238.

%Acknowledgements are not compulsory. Where included they should be brief. Grant or contribution numbers may be acknowledged.

%Please refer to Journal-level guidance for any specific requirements.

\bmhead{Author contributions}

I.N.U.\ performed the surface SPM experiments and prepared the samples and K.A.H.\ assisted for the initial experiments. M.T.\ prepared wedge specimens for STEM supervised by A.T.J.H.; J.H.\ prepared FIB-cut lamellas for STEM. M.T.\ and U.L.\ carried out STEM measurements and performed corresponding data analysis with support from A.T.J.H.; J.I-G.\ performed DFT calculations and symmetry analysis. M.Z.K.\ calculated electromechanical tensors using DFT supervised by S.M.S.; M.S-M.\ performed FEM simulations under supervision of J.G.; N.S.\ performed T-AFM and analyzed the data under supervision of A.K.; Ch.G.\ performed P(E) measurements supervised by J.H.; P.B.\ and L.B.\ suggested and provided the materials and contributed to discussions on crystallographic properties. The work of I.N.U., J.H., U.L., and K.A.H.\ was done under the supervision of D.M.; D.M.\ devised and coordinated the project. I.N.U.\ and D.M.\ wrote the manuscript, with support from J.I-G.\ for the DFT- and symmetry-related sections. All authors discussed the results and contributed to the final version of the manuscript.

\pagebreak

\includepdf[pages=-]{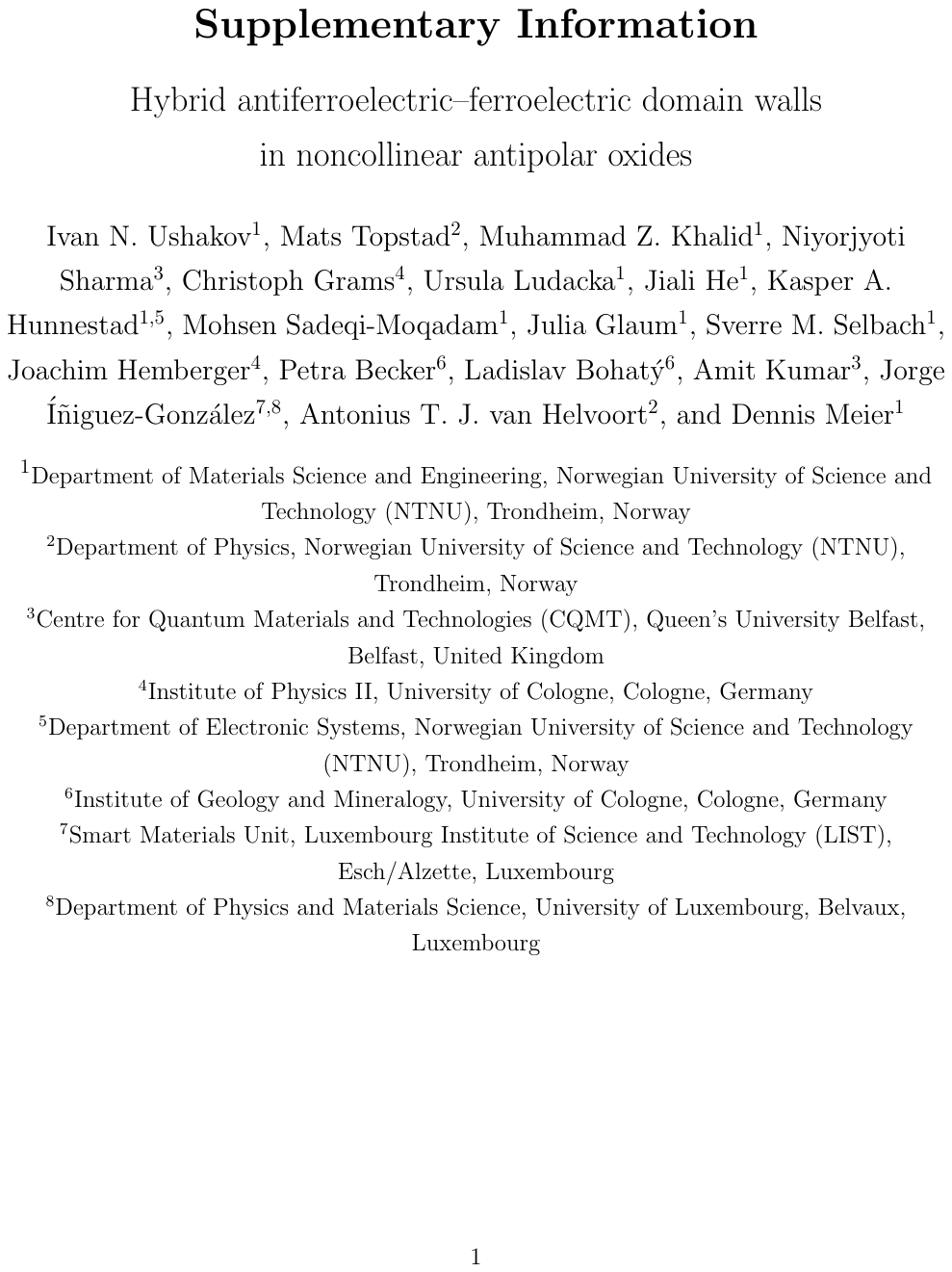}

%\bibliography{sup1-bibliography}% common bib file

\end{document}